\documentclass[twocolumn,showpacs,superscriptaddress,preprintnumbers,amsmath,amssymb]{revtex4-1}

\usepackage{graphicx,tabularx}  
\usepackage[english]{babel}
\usepackage{dcolumn}            
\usepackage{bm}                 
\usepackage{array}              
\usepackage{hhline}             
\usepackage{multirow}
\usepackage{color}
\usepackage{upgreek}          
\usepackage{xtdeqnra} 

\begin{document}

\preprint{APS/123-QED}

\title{Enhancing magnetic near--field intensities with dielectric resonators}

\author{Guillaume Boudarham}
\affiliation{CNRS, Aix-Marseille Universit\'e, Centrale Marseille, Institut Fresnel, UMR 7249, 13013 Marseille, France}
\author{Redha Abdeddaim}
\affiliation{CNRS, Aix-Marseille Universit\'e, Centrale Marseille, Institut Fresnel, UMR 7249, 13013 Marseille, France}
\author{Nicolas Bonod}
\email{nicolas.bonod@fresnel.fr}
\affiliation{CNRS, Aix-Marseille Universit\'e, Centrale Marseille, Institut Fresnel, UMR 7249, 13013 Marseille, France}

\date{\today}

\begin{abstract}
We measure strong magnetic field intensities in a subwavelength gap separating two dielectric resonators. 
This dimer magnetic antenna is characterized in the GHz spectral regime when considering three conditions of illumination. We detail the different magneto-electric couplings involved in the enhancement of the near magnetic field and derive the analytical expression of the magnetic field intensities. The results reported here in the GHz domain can be extended to other spectral domains since they are obtained with a dielectric permittivity 16 that can be observed in a very broad spectrum ranging from optics to radio-frequencies.
\end{abstract}

\maketitle 
Structuring matter at a scale smaller than the wavelength has allowed to create artificial meta-atoms that exhibit large electric and 
magnetic polarization moments when excited by an electromagnetic field. Metallic split ring or U-shaped resonators \cite{Husnik08,Burresi09,Sersic11,Varault13,Kwadrin13,Hein13}, photonic crystal cavities \cite{Burresi10,Vignolini10}, nanostructured dielectric surfaces \cite{Girard97,Devaux00} or also inverse antennas \cite{Popov07,Grosjean11,Kihm11} proved their ability to enhance local magnetic fields. High dielectric permittivity resonators supporting Mie resonances can also resonantly interact with incident electromagnetic 
field via electric or magnetic photonic modes \cite{OBrien02,Schuller07,Popa08,Zhao09}. 
More recently, it was evidenced that single semi-conductor particles can offer electric and 
magnetic Mie resonance in the near-infrared and optical frequencies \cite{Evlyukhin10,Garcia-Etxarri11,Evlyukhin12,Kuznetsov12}.  
These resonators could open the route towards the design of magneto-electric antennas \cite{Rolly12b,Schmidt12,Albella13} to control the electric and magnetic transitions of emitters such as trivalent lanthanide ions \cite{Thommen06,Noginova09,Karaveli11,Dodson12,Taminiau12} or to enhance circular dichroism \cite{Dionne13}.

\begin{figure}[t!]
\centerline{\includegraphics[scale=0.2]{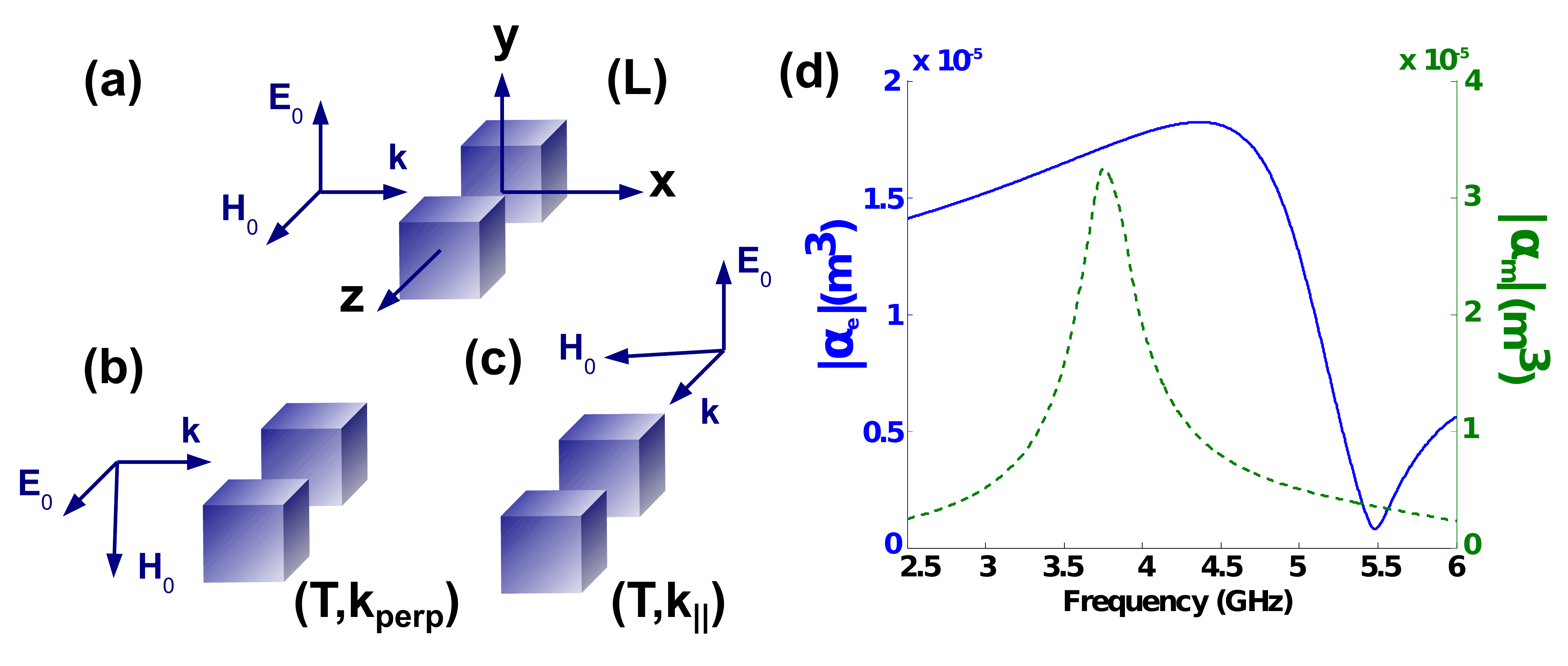}}
\caption{(Color online) At left : dimer illumination schematics for longitudinal (a) and transverse (b-c) illuminations. The origin of the coordinate system is chosen in the center of the gap of the dimer and the $z$-axis is chosen to be aligned with the dimer axis. 
At right : Modulus of the electric (blue line, left scale) and magnetic (green dashed line, right scale) polarizability of an individual cube as a function of the frequency.}
\label{figure1}
\end{figure}

Here we investigate the near-field properties of magneto-dielectric scatterers. We present an experiment set-up able to measure the enhancement of the magnetic field intensity inside the 
gap of a dimer of Mie resonators. The dielectric antenna is illuminated by a plane wave under three different illuminations 
(see Figs.~\ref{figure1} (a-c)) for which
the incident wavevector $\bm{k}$ can be parallel (illumination $T,k_{||}$) or normal to the dimer axis. In the latter case, the incident magnetic field $\bm{H}_0$ can be either parallel (illumination L) or normal to the dimer axis (illumination $T,k_\bot$) :
\begin{eqnarray}
\bm{H}_{0,L}(x) &=& |\bm{H}_0|e^{ikx}\hat{\bm{e}}_{z},  \\
\bm{H}_{0,T,k_\bot}(x) &=& -|\bm{H}_0|e^{ikx}\hat{\bm{e}}_{y}, \\
\bm{H}_{0,T,k_\parallel}(z) &=& -|\bm{H}_0|e^{ikz}\hat{\bm{e}}_{x}.  
\end{eqnarray}
The experimental measurements are carried out in the GHz regime with resonators exhibiting a dielectric permittivity of $16+i0.25$ (Eccostock HIK, K=16) and a size length of $16$ mm. The measurements are performed in an anechoic chamber with a network analyzer (IF filter was calibrated at 10 Hz). The samples are positioned on an expanded polystyrene substrate which is almost equivalent to air at our working frequencies and are illuminated with a broadband horn antenna (1-18\,GHz). 
This antenna is used to maintain a linear polarization in the entire frequency range. The magnetic field is measured by a small near field probe (MFA03 by LANGER).
The frequency range of the probe is 0.2-6\,GHz with a spatial resolution of 200\,$\mu$m. 
The level of cross-polarization is less than 20\,dB that ensures a good decoupling between the different components of the measured field. 

We first characterize a single dielectric cube resonator illuminated 
by a plane wave $(\bm{E}_0,\bm{H}_0)$ propagating along the $x$-axis, where $\bm{E}_0||\hat{\bm{y}}$ and $\bm{H}_0||\hat{\bm{z}}$. Fig.~\ref{figure1} at right shows the modulus of the
electric and magnetic polarizability of the cube as a function of the incidence frequency and obtained from a finite element method (FEM) analysis. We can clearly observe a magnetic dipolar resonance at 3.75\,GHz.
We measured the three components of the magnetic field intensity by placing the probe at 1\,mm from the center of the cube face along the $Oz$ axis and oriented to be sensitive either to the $x$, $y$ or $z$-component of the magnetic field. When the incident magnetic field is along the $z$-axis, we found that the normalized
magnetic field intensity $|\bm{H}_z|^2/|\bm{H}_0|^2$ is maximal at 3.69\,GHz where it reaches 60. The normalization is carried out with respect to the magnetic field measured at the same position without cube. 
Accordingly to Fig.~\ref{figure1} (d), this enhancement results from the excitation of the magnetic dipolar mode inside the cube. The two other components which are normal to the incident magnetic field were measured at a level below the dynamic range of the network analyser (-90\,dB), meaning that the total field $\bm{H}$ can be identified here to $\bm{H}_z$. 

\begin{figure*}[t!]
\centerline{\includegraphics[scale=0.32]{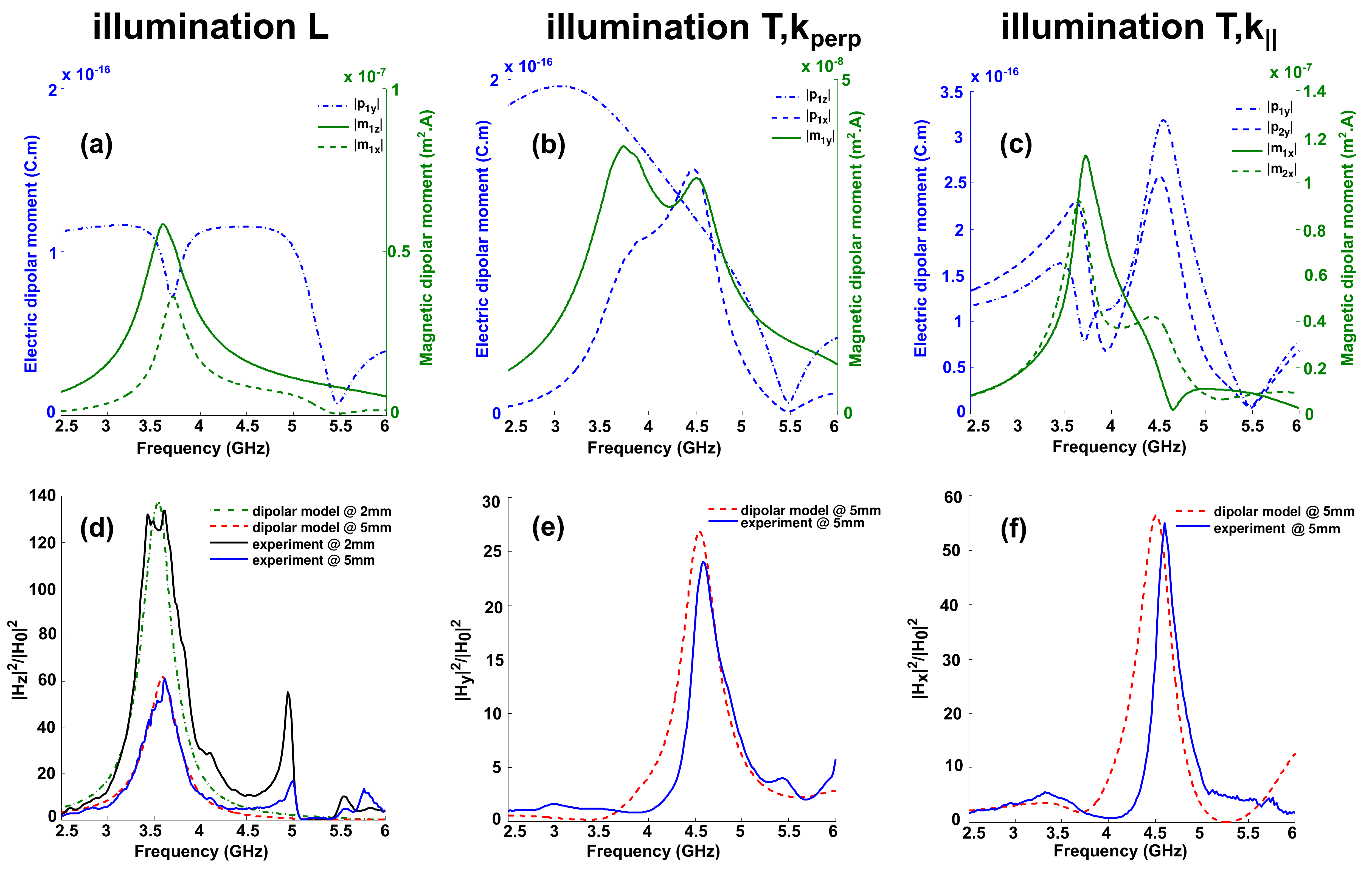}}
\caption{(Color online) (a)-(c): modulus of the dipolar moments obtained from the dipolar model for a gap size of 5\,mm for the three conditions of illumination.  
(d)-(f) : normalized magnetic field intensity obtained from the dipolar model (dashed lines) and experimental measurements (full lines), for a gap size of 5\,mm (d-f) or 2\,mm (d).}
\label{figure2}
\end{figure*}

We derive a semi-analytical model to calculate the magnetic field intensity inside the 
gap of the dimer for the three illuminations. We use the international system of units and we will consider $|\bm{E}_0|=1 (V/m)$. A particle $n$ ($n=1,2$) of the dimer located at $\bm{r}_n$ 
is assumed to respond with an induced electric and magnetic dipole respectively noted $\bm{p}_n$ and $\bm{m}_n$. 
The dipoles of each particle, that we assume here identical, non-bianisotropic and described by the polarizability tensor 
$\bar{\bar{\bm{\alpha}}}=\text{diag}(\bar{\bar{\bm{\alpha}}}_e,\bar{\bar{\bm{\alpha}}}_m)$, where $\bar{\bar{\bm{\alpha}}}_e$ and 
$\bar{\bar{\bm{\alpha}}}_m$ respectively denote the electric and magnetic polarizability tensors of the individual particles, 
verify the general tensorial system of equations \cite{de2007colloquium,Albella13} :

\begin{equation}\label{syst1_prl}
\left(\begin{array}{c}
\bm{p}_n/\epsilon_0 \\
\bm{m}_n 
\end{array}
\right)=\bar{\bar{\bm{\alpha}}}\cdot\left[\left(\begin{array}{c}
\bm{E}_0(\bm{r}_n) \\
\bm{H}_0(\bm{r}_n) 
\end{array}
\right)+\sum\limits_{\substack{q=1,2 \\ q \neq n}}\bar{\bar{\bm{G}}}^0(\bm{r}_n,\bm{r}_q)\cdot\left(\begin{array}{c}
\bm{p}_q \\
\bm{m}_q 
\end{array}
\right)\right],
\end{equation}

where 
\begin{equation}
\bar{\bar{\bm{G}}}^0(\bm{r},\bm{r}')=\displaystyle\left(
\begin{array}{cc}
\bar{\bar{\bm{G}}}^0_{ee}(\bm{r},\bm{r}') & \bar{\bar{\bm{G}}}^0_{em}(\bm{r},\bm{r}') \\
\bar{\bar{\bm{G}}}^0_{me}(\bm{r},\bm{r}') & \bar{\bar{\bm{G}}}^0_{mm}(\bm{r},\bm{r}') 
\end{array}
\right)
\end{equation}
is the full dyadic Green's function of free space \cite{Sersic11}. 
The electric and magnetic polarizability tensors of the individual particles are diagonal and can be described by 
scalars : $\bar{\bar{\bm{\alpha}}}_e=\alpha_e\text{diag}(1,1,1)$ and $\bar{\bar{\bm{\alpha}}}_m=\alpha_m\text{diag}(1,1,1)$,
where $\alpha_e\equiv\alpha_{ee}^{yy}=p_y/\epsilon_0$ and $\alpha_m\equiv\alpha_{mm}^{zz}=m_z Z_0$,
where $Z_0$ is the impedance of free space \cite{jackson1999}. $\alpha_e$ and $\alpha_m$ can be deduced from the knowledge of the electric and magnetic dipolar moments of the individual cube $\bm{p}$ and $\bm{m}$ that are obtained by calculating the polarization density $\bm{P}$ with a FEM analysis: $\bm{p}=\int\bm{P}d^3\bm{r}$ and 
$\bm{m}=-j\omega/2\int\bm{r}\times\bm{P}d^3\bm{r}$, where the integration is performed over the volume of the structure \cite{jackson1999}. \\

To obtain $\bm{p}_1$, $\bm{p}_2$, $\bm{m}_1$ and $\bm{m}_2$ for any incidence and polarization of the incident field, we define :
\begin{equation}
\bm{D} \equiv \left(
\begin{array}{c}
\bm{p}_1 \\
\bm{p}_2 \\
\bm{m}_1 \\
\bm{m}_2
\end{array}
\right)~~\text{and}~~\bm{F} \equiv \left(
\begin{array}{c}
\epsilon_0\alpha_e\bm{E}_0(\bm{r}_1) \\
\epsilon_0\alpha_e\bm{E}_0(\bm{r}_2) \\
\alpha_m\bm{H}_0(\bm{r}_1) \\
\alpha_m\bm{H}_0(\bm{r}_2) 
\end{array}
\right),
\end{equation}
to cast equations (\ref{syst1_prl})  in the form : $\bar{\bar{\bm{M}}}.\bm{D} = \bm{F}$ where:

\begin{widetext}
\begin{equation}\label{matrix}
\bar{\bar{\bm{M}}} \equiv \left(
\begin{array}{cccc}
\bar{\bar{\bm{I}}} & -\epsilon_0\alpha_e\bar{\bar{\bm{G}}}^0_{ee}(\bm{r}_1,\bm{r}_2) & \bm{0} & -\epsilon_0\alpha_e\bar{\bar{\bm{G}}}^0_{em}(\bm{r}_1,\bm{r}_2) \\
-\epsilon_0\alpha_e\bar{\bar{\bm{G}}}^0_{ee}(\bm{r}_2,\bm{r}_1) & \bar{\bar{\bm{I}}} & -\epsilon_0\alpha_e\bar{\bar{\bm{G}}}^0_{em}(\bm{r}_2,\bm{r}_1) & \bm{0} \\
\bm{0} & -\alpha_m\bar{\bar{\bm{G}}}^0_{me}(\bm{r}_1,\bm{r}_2) & \bar{\bar{\bm{I}}} & -\alpha_m\bar{\bar{\bm{G}}}^0_{mm}(\bm{r}_1,\bm{r}_2) \\
-\alpha_m\bar{\bar{\bm{G}}}^0_{me}(\bm{r}_2,\bm{r}_1) & \bm{0} & -\alpha_m\bar{\bar{\bm{G}}}^0_{mm}(\bm{r}_2,\bm{r}_1) & \bar{\bar{\bm{I}}}
\end{array}
\right).
\end{equation}
\end{widetext}

$\bar{\bar{\bm{M}}}$ is a $12\times 12$ block matrix which completely describes the dimer regardless of the incident field.
$\bar{\bar{\bm{I}}} = \text{diag}(1,1,1)$ and $\bm{0}$ respectively denote the $3\times 3$ unit dyadic and the dyadic filled with zero. \\

The electric and magnetic dipolar moments of each particle can thus be obtained by inverting the matrix 
$\bar{\bar{\bm{M}}}$ : $\bm{D}=\bar{\bar{\bm{M}}}^{-1}.\bm{F}$. The explicit expressions of the moments and useful discussions are detailed 
in the Supplemental Material. They are plotted in Figs.~\ref{figure2} (a-c) as a function of the frequency for a gap size of 5\,mm. \\

The latter formulas allow to derive the expression of the magnetic field in the center of the gap for the three conditions of 
illumination \cite{jackson1999}:

\begin{widetext}
\begin{eqnarray}
H_{z,L} & = & \frac{e^{ikl}}{2\pi l^3}(1-ikl) m_{1z} + \frac{e^{ikl}}{2\pi l^3}(1-ikl) m_{2z}, \label{Hz}  \\
H_{y,T,k_\perp} & = &\frac{e^{ikl}}{4\pi l}k\omega\left(1+\frac{i}{kl}\right)p_{1x}-\frac{e^{ikl}}{4\pi l}k\omega\left(1+\frac{i}{kl}\right)p_{2x}
    + \frac{e^{ikl}}{4\pi l^3}(k^2 l^2+ikl-1)m_{1y}+\frac{e^{ikl}}{4\pi l^3}(k^2 l^2+ikl-1)m_{2y}, \label{Hy} \\
H_{x,T,k_\parallel} & = & -\frac{e^{ikl}}{4\pi l}k\omega\left(1+\frac{i}{kl}\right)p_{1y} + \frac{e^{ikl}}{4\pi l}k\omega\left(1+\frac{i}{kl}\right)p_{2y} 
    + \frac{e^{ikl}}{4\pi l^3}(k^2l^2+ikl-1) m_{1x} + \frac{e^{ikl}}{4\pi l^3}(k^2l^2+ikl-1) m_{2x} \label{Hx}
\end{eqnarray}
\end{widetext}

where $l$ is the distance between the center of a cube and the origin of the coordinate system. The total magnetic field is obtained by adding to the previous fields the incident magnetic field. We have considered only the component of the magnetic field in the direction of the incident magnetic field because the other components reveal to be negligible :\\   
Due to the finite size of the probe, the distance between the particles of the dimer will be necessarily larger than 4\,mm, except with the L-illumination (Fig.1(a)). 
One can see in Fig.~\ref{figure2} (d) (red dashed line or blue line) that the magnetic field intensity is maximum at 3.52\,GHz where it reaches 61. This maximum coincides with the resonance of the $z$-component of the magnetic dipolar moment 
[see Fig.~\ref{figure2} (a), green line]. The second peak experimentally observed at 5.0\,GHz (blue line) is not 
reproduced by the dipolar model and is due to the excitation of the magnetic quadrupolar resonance. 
The experimental measurements associated to the dipolar model demonstrates that for this illumination condition, the strong magnetic field enhancement results from an efficient coupling between two magnetic dipolar modes 
aligned with the dimer axis ($z$-axis). In Fig.~\ref{figure2} (e) (blue line), the maximum of the normalized magnetic field intensity is measured at 4.58\,GHz where it reaches 24, which is almost three times lower than the enhancement previously reported. This maximum coincides with both a maximum of the $y$-component 
of the magnetic dipolar moment and a maximum of the $x$-component of the electric dipolar moment (see Fig.~\ref{figure2} (b)). 
One can see in Fig.~\ref{figure2} (f) (blue line) that the normalized magnetic field intensity is maximum at 4.60\,GHz where it reaches 55. 
In this case, the peak matches a maximum of the $y$-component of the electric dipolar moments of each particle and the $x$-component of
the magnetic dipolar moment of the second particle (Fig.~\ref{figure2} (c)). 
The dipolar model shows a remarkable 
agreement with experiments especially for the longitudinal illumination and the principal resonance for the transverse 
illuminations. This good concordance between the dipolar model and experiments is due to the weak sensitivity of the dipolar resonances to high spatial frequency of the resonator shapes. This means that smoothing the edges of the resonators will not strongly impact the magnetic field enhancement mostly related here to the dipolar polarization moments. 
We also studied the influence of the gap size on the magnetic field intensity. In the case of a longitudinal illumination (L), we were able to reduce the gap size down to 2\,mm and measured enhancement as high as 130 (Fig.~\ref{figure2} (d), black line). The analytical model was applied to verify that a decrease of the gap also leads to an increase of $|\bm{H}|^2/|\bm{H}_0|^2$ for the both transverse illuminations. In a sake of completeness, we estimated with both complete calculations and analytical model the electric field intensity and observed that its enhancement is limited to 20 (occurring for illumination L), and remains in general small for the three illuminations. This result is important since it proves that the magnetic field can be enhanced independently from the electric field 
(see supporting information). \\
To summarize, strong magnetic field intensities are measured when coupling two identical dielectric resonators illuminated upon three different illuminations. The largest enhancement results from a strong coupling between the magnetic modes due to the efficient coupling of the magnetic dipolar 
moments which are almost parallel to the dimer axis. Magnetic field intensities enhanced by more than two orders of magnitude were measured. Experimental investigations were completed by a detailed dipolar analytical model that was able to predict the magnetic and electric field intensity in the vicinity of a magneto-dielectric dimer illuminated by a 
plane wave in arbitrary incidence and polarization. Dimers of plasmonic resonators have been widely investigated to enhance at a subwavelength scale the electric light matter interaction \cite{Novotny11,Busson12}, we believe dielectric magnetic antennas will open the way towards light matter interaction enhanced $via$ the magnetic field, with important applications in enhanced spectroscopy and resonant light cavities.

\begin{acknowledgments}
The authors would like to thank Elodie Georget for preliminary experiments, Brice Rolly for his support in numerical simulations, Jean Michel Geffrin and Sebastien Bidault for fruitful discussions.
This research was funded by the French Agence Nationale de la Recherche under Contract No. ANR-11-BS10-002-02 TWINS.
\end{acknowledgments}

\end{document}